\documentclass{article}
\usepackage{spconf,amsmath,graphicx}

\usepackage{enumitem}
\setlist{nosep, leftmargin=14pt}
\let\svthefootnote\thefootnote
\newcommand\freefootnote[1]{%
  \let\thefootnote\relax%
  \footnotetext{#1}%
  \let\thefootnote\svthefootnote%
}
\usepackage[utf8]{inputenc}
\usepackage[english]{babel}
\usepackage{cite}
\usepackage{graphicx}

\usepackage{amsfonts} 
\usepackage{float}
\usepackage{xcolor}
\usepackage{caption}
\usepackage{subcaption}
\newcommand{\cmmnt}[1]{\ignorespaces}
\usepackage{eqnarray,amsmath}
\usepackage{amsmath,bm}
\usepackage{amssymb}
\usepackage{graphicx}
\usepackage[linesnumbered,ruled,vlined]{algorithm2e}

\title{ Regularization by Adversarial Learning for Ultrasound Elasticity Imaging}
\name{%
    Narges Mohammadi$^{\star}$%$^{\star \dagger}$%
    \qquad Marvin M. Doyley$^{\star}$%
    \qquad Mujdat Cetin$^{\star}{}^{\dagger}$}
\address{$^{\star}$ Department of Electrical and Computer Engineering, University of Rochester, Rochester, NY, USA \\%
   $^{\dagger}$ Goergen Institute for Data Science, 
University of Rochester, Rochester, NY, USA
}

\begin{document}
%\ninept
\maketitle
\begin{abstract}
Classical model-based imaging methods for ultrasound elasticity inverse problem \cmmnt{reconstruction} require prior constraints about the underlying elasticity patterns, while finding the appropriate hand-crafted prior for each tissue type is a challenge.
In contrast, standard data-driven methods count solely on supervised learning on the training data pairs leading to massive network parameters for unnecessary physical model relearning which might not be consistent with the governing physical models of the imaging system. Fusing the physical forward model and noise statistics with data-adaptive priors leads to a united reconstruction framework that guarantees the learned reconstruction agrees with the physical models while coping with the limited training data. In this paper, we propose a new methodology for estimating the elasticity image by solving a regularized optimization problem which benefits from the physics-based modeling via a data-fidelity term and adversarially learned priors via a regularization term. In this method, the regularizer is trained based on the Wasserstein Generative Adversarial Network (WGAN) objective function which tries to distinguish the distribution of clean and noisy images. Leveraging such an adversarial regularizer for parameterizing the distribution of latent images and using gradient descent (GD) for solving the corresponding regularized optimization task leads to stability and convergence of the reconstruction compared to pixel-wise supervised learning schemes. Our simulation results verify the effectiveness and robustness of the proposed methodology with limited training datasets.
\end{abstract}
\cmmnt{which can be extended to unsupervised training 
 In this method, physics-based modeling is performed using a statistical linear representation of equilibrium equations unveiling the relationship of measured data to the unknown elasticity modulus. Additionally, the adversarial regularizer is exploited based on a Wasserstain Generative Network (WGAN) to express the underlying prior structure of the tissue.}
\vspace{-0.45cm}
\section{Introduction}
\vspace{-0.2cm}
\freefootnote{This work has been partially supported by the National Science Foundation (NSF) under Grants CCF-1934962 and DGE-1922591.}
Ultrasound elastography as a non-invasive tool for tissue stiffness characterization offers great potential for reliable clinical diagnosis. Elasticity image reconstruction can be accomplished by solving an inverse problem formulated as a constrained optimization problem under regularization. This optimization task consists of a forward model describing the physics of the imaging system and a regularization term expressing any prior information about the latent image. Considering medical imaging challenges including time-efficient image reconstruction as well as reliable and robust reconstructed elastic images, numerous advances have been proposed. One major concern in existing model-based approaches \cite{Marvin}, \cite{narges2} is how to capture the appropriate prior information about the complex structure of the underlying tissues and how to incorporate this prior knowledge into the image reconstruction scheme. By the advent of deep neural networks (DNNs) \cite{salman2}, various end-to-end learning-based methods \cite{E2E, E2E2, E2E3, tehrani, mahsa1} have been proposed which try to learn both the physical model and the prior information about the underlying tissues. These methods lead to many shortcomings such as very large number of training pairs requirements and no-guaranteed solutions consistent with the true physical models. Moreover, these methods prevent a high level of generalizability which means that expensive network retraining is required whenever the forward model, noise distribution, or noise level changes. These limitations have been overcome by combining forward models and learned priors in a constrained optimization task, resulting in more accurate and time-efficient imaging schemes \cite{willet,PGD,selfsuper, ali}.
Such approaches can be split into two groups: unrolling-based methods and prior learning methods.\\
Unrolling-based approaches integrate physical models into the learning process by unfolding each iteration of the classical optimization problem as a layer of a neural network which includes algorithms such as PINN \cite{PINN}, PI-GAN \cite{GAN}, \cite{Moji} and \cite{salman1}. These approaches provide improved accuracy while they are time-consuming as they require network retraining for each iteration.\\
On the other hand, prior learning methods infuse \cmmnt{(bringing)} learned priors as regularizers into physics-model based inversion \cmmnt{(try to have learned units embedded in model-based image reconstruction)} which includes two types of methods.
First, supervised learning of regularizers using training pairs is utilized in methods such as Plug-and-Play (PnP) \cite{pnp} and regularization by denoising (RED) \cite{RED}. PnP methods use pixel-wise loss functions for learning a denoiser which then acts as an implicit regularizer within an iterative reconstruction approach such as alternating direction method of multipliers (ADMM) and proximal gradient methods \cite{primaldual}. Second, unsupervised learning of regularizers using an adversarial network can be used to construct explicit regularizers that can be incorporated into the optimization problem for imaging. These learning-based priors estimate the distribution of the latent image using generative adversarial networks (GAN) \cite{GANprior}, Wasserstein GAN (WGAN) \cite{WGAN} or variational auto-encoders (VAE). WGAN is an augmented GAN that replaces the discriminator (which predicts the probability of fakeness or realness of generated images) with a critic which tries to score the fakeness or realness of the given images leading to a better approximation of the distribution of the observed data in the training dataset \cite{morteza2}. Regarding some advantages of WGAN over GAN, we can refer to more stability during network training and less sensitivity to network design and hyperparameter settings. Since adversarial regularizers are trained based on image distribution loss, rather than image pixel loss, no paired training data is necessary. This, in principle, allows for some level of unsupervised training. In addition, the ability to formulate the imaging problem with explicit priors provides some interpretability. Finally, we can also benefit from the theoretical stability and convergence results on the resulting optimization problems for imaging \cite{carola}.  \cmmnt{Moreover, the critic loss function implies the reconstructed image quality.}\\
In this article, we propose a new elasticity image reconstruction framework that incorporates an explicit adversarially learned regularizer into an optimization formulation that also involves a physics-based forward model and a noise model. In this approach, the image prior is learned using an adversarial critic network based on the WGAN objective function which seeks to discriminate between the distributions of regularization-free reconstructions and the ground-truth images. Once the regularizer is trained, it is plugged into the constrained optimization task for solving the inverse problem using a gradient descent scheme. \cmmnt{for optimizing its linear algebraic operations. As we are using probabilistic modeling, we are able to efficiently benefit from noise and error modeling.}
Our simulation results verify the effectiveness of the proposed methodology for elasticity image reconstruction with limited training datasets and noisy displacement fields.\\
The remainder of this paper is organized as follows. In Section 2, we describe the constrained optimization problem for ultrasound elasticity imaging. The proposed adversarial learning-based elasticity imaging approach is described in Section 3. Section 4 presents the results of our preliminary experimental analysis, and finally, concluding remarks are available in Section 5.
\vspace{-0.25cm}
\section{Optimization Problem Formulation}
\vspace{-0.15cm}
\cmmnt{For data-fidelity term, we model the equilibrium equation of elasticity as a linear regression model with respect to elasticity modulus which leads to a weighted least square term considering the covariance matrix as a signal-dependent correlated one described in our previous work.\\}
The imaging system for ultrasound elastography of incompressible tissues can be modeled using the quasi-static equilibrium equation. This model also known as the global stiffness equation reveals the relationship between the unknown elasticity of tissue $\mathbf{x}$ with the observed deformation measurements $\mathbf{u}$ in response to the applied force measurements $\mathbf{f}$ as follows:
\vspace{-0.5cm}
\begin{equation}
\label{eq:1}
\mathbf{f}=\mathbf{K}(\mathbf{x})\mathbf{u}+\mathbf{w}\qquad \mathbf{w}\sim \mathcal{N}(0,\,\bm{\Sigma_{w}})
\vspace{-0.1cm}
\end{equation}
Utilizing the finite-element-method (FEM) for medium discretization over $N$ nodes of a mesh, $\mathbf{f}\in \mathbb{R}^{2N\times 1}$ stands for the nodal force measurements constituting the medium boundary condition, $\mathbf{u}\in \mathbb{R}^{2N\times 1}$ represents the true nodal displacements in the lateral and axial dimensions and $\mathbf{w}\in \mathbb{R}^{2N\times 1}$ expresses the nodal Gaussian noise. The main role of elasticity $\mathbf{x}\in \mathbb{R}^{N\times 1}$ as the mechanical tissue characteristic is embedded in $\mathbf{K}(\mathbf{x})\in  \mathbb{R}^{2N\times 2N}$ which describes the force and deformation field relationship.
Estimating the elastic modulus $\mathbf{x}$ as an inverse problem can be fulfilled by solving a constrained optimization problem. In this regard, the forward model (\ref{eq:1}) needs to be formulated as a linear representation with respect to the unknown elasticity modulus \cite{narges2}. To this end, we establish the matrix $\mathbf{D}(\mathbf{u})\in \mathbb{R}^{2N\times N}$ which has the following relation with $\mathbf{K}(\mathbf{x})$ using a 3D tensor $\bm{\Psi}\in \mathbb{R}^{N\times 2N\times2N}$ constructed from the equilibrium equations:
\begin{equation}
\begin{array}{l}
\label{eq:2}
\qquad\quad\mathbf{D}(\mathbf{u})\mathbf{x}=\mathbf{K}(\mathbf{x})\mathbf{u}\\
\mathbf{D}(\mathbf{u})=(\bm{\Psi}\mathbf{u})^{T} \qquad \mathbf{K}(\mathbf{x})=\bm{\Psi}^{T}\mathbf{x}
\end{array}
\vspace{-0.1cm}
\end{equation}
In practice, the deformation fields are measured by cross-correlation of several B-mode ultrasound images which leads to the noisy deformation field measurements $\mathbf{u^{m}}=\mathbf{u}+\mathbf{n}$ where $\mathbf{n}\sim \mathcal{N}(0,\,\bm{\Sigma_{n}})$. Substituting this relationship into the forward model (\ref{eq:1}) yields:
\vspace{-0.1cm}
\begin{eqnarray}
\label{eq:5}
\mathbf{f}&=&\mathbf{K}(\mathbf{x})\mathbf{u}+\mathbf{w}=\mathbf{K}(\mathbf{x})(\mathbf{u^{m}}-\mathbf{n})+\mathbf{w}\nonumber\\
&=&\mathbf{K}(\mathbf{x})\mathbf{u^{m}}-\mathbf{K}(\mathbf{x})\mathbf{n}+\mathbf{w}
\end{eqnarray}
Letting ${\mathbf{\Tilde{w}}}=-\mathbf{K}(\mathbf{x})\mathbf{n}+\mathbf{w}$ and utilizing (\ref{eq:2}) with noisy deformations, $\mathbf{D}(\mathbf{u^{m}})\mathbf{x}=\mathbf{K}(\mathbf{x})\mathbf{u^{m}}$, the integrated forward model can be represented as:
\vspace{-0.2cm}
\begin{equation}
\label{eq:6}
\mathbf{f}=\mathbf{D}(\mathbf{u^{m}})\mathbf{x}+\mathbf{\Tilde{w}}\qquad \mathbf{\Tilde{w}}\sim \mathcal{N}(0,\,\bm{\Gamma})
\end{equation}
where $\bm{\Gamma}$ is given by:
\vspace{-0.1cm}
\begin{equation}
\label{eq:7}
\bm{\Gamma}=\bm{\Sigma_{w}}+\mathbf{K}(\mathbf{x})\bm{\Sigma_{n}}\mathbf{K}(\mathbf{x})^{T}
\vspace{-0.15cm}
\end{equation}
based on which we can interpret (\ref{eq:6}) as a linear observation model involving signal dependent colored noise. This statistical forward model which incorporates the noise statistics paves the way for formulating the elastic inverse problem using a regularized optimization problem \cite{narges2} as follows:
\begin{equation}
\label{eq:8}
\begin{array}{l}
\hat{\mathbf{x}}=\mathrm{argmin} _{\mathbf{x}}\quad\frac{1}{2}\left \|  \mathbf{f}-\mathbf{D}(\mathbf{u^{m}})\mathbf{x} \right \|_{{\bm{\Gamma}}^{-1}}^{2}+\lambda R(\mathbf{x})\\
\quad\quad\quad
s.t.\quad \mathbf{x}>0
\end{array}
\vspace{-0.25cm}
\end{equation}
where $\left \| \mathbf{A} \right \|_{\mathbf{B}}^{2}:=(\mathbf{A}^{T}\mathbf{B}\mathbf{A})$ and $R(\mathbf{x})$ is the regularization term. To leverage the potential of learning-based methods in capturing prior information about the underlying scenes, we learn the regularization term using an adversarial critic network $\mathbf{C}_{w}(\mathbf{x})$ with learned weights $w$ using ground-truth elasticity modulus $\mathbf{x}$ as inputs.\\
\begin{figure*}[t!]%[t]
  \centering
  \centerline{\includegraphics[width=14cm]{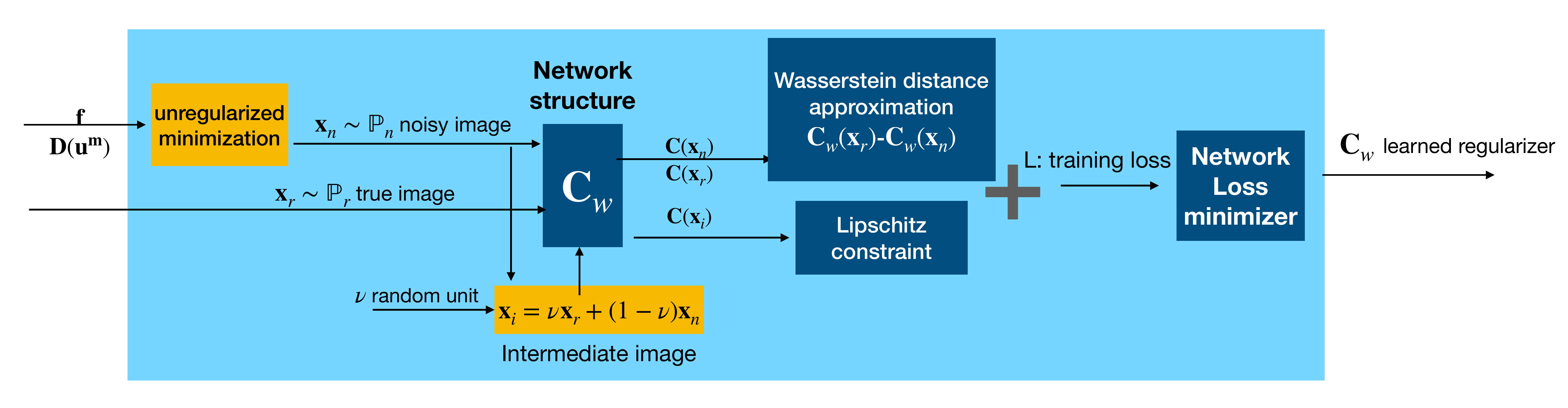}}
  \vspace{-0.35cm}
\caption{Training the adversarial regularizer using critic network optimization.}
\label{fig:1}
\vspace{-0.35cm}
\end{figure*}
\begin{figure*}[ht]%[t]
  \centering
  \centerline{\includegraphics[width=10cm]{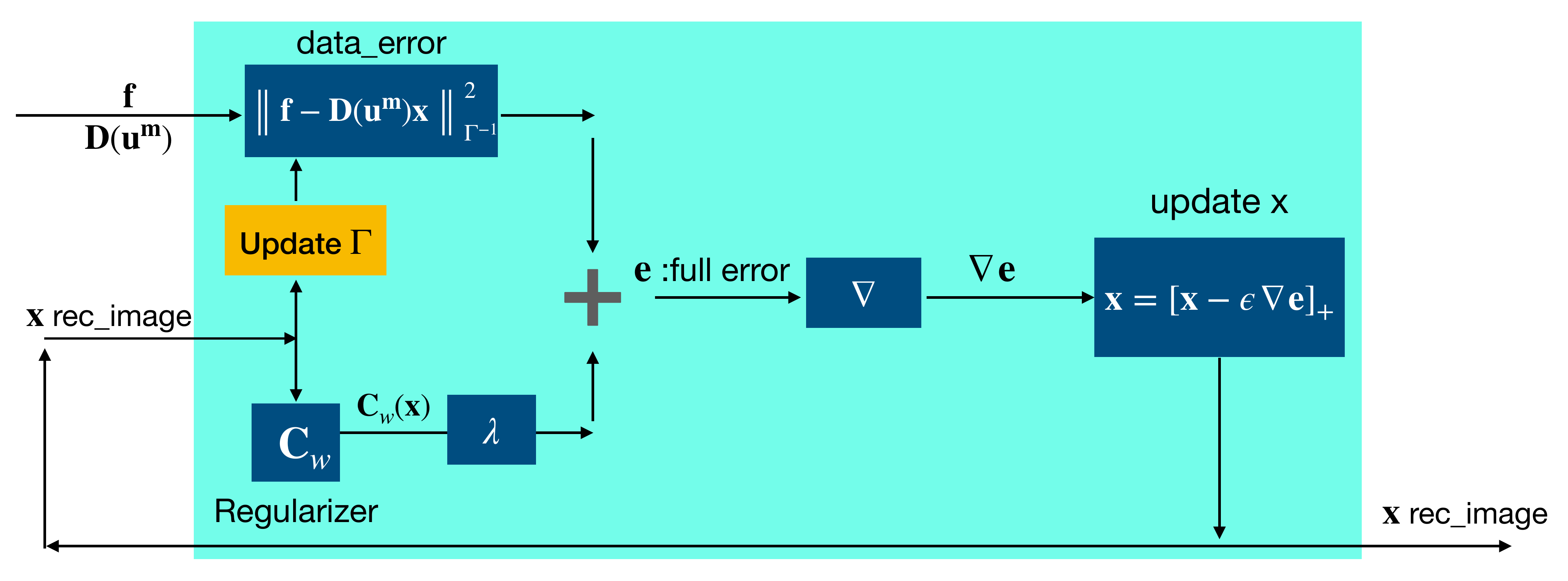}}
  \vspace{-0.35cm}
%  \centerline{(a)}\medskip
\caption{Elasticity image reconstruction using adversarially learned regularization.}
\label{fig:2}
\end{figure*}
After obtaining the learned regularizer $\mathbf{C}_{w}(\mathbf{x})$, we use a fixed-point approach to solve (\ref{eq:8}), where $\bm{\Gamma}$ will be fixed when updating $\mathbf{x}$, and then $\bm{\Gamma}$ will be updated with the new $\mathbf{x}$ based on (\ref{eq:7}).
For updating $\mathbf{x}$ in each step of the fixed-point approach, we use gradient descent (GD): %\cite{proximal}:
\begin{equation}
\label{eq:9}
\mathbf{x}\xleftarrow{}[\mathbf{x}-\epsilon \nabla_{\mathbf{x}} (\left \|  \mathbf{f}-\mathbf{D}(\mathbf{u^{m}})\mathbf{x} \right \|_{{\bm{\Gamma}}^{-1}}^{2}+\lambda \mathbf{C}_{w}(\mathbf{x}))]_{+}
\end{equation}
where $[]_{+}$ stands for the positivity constraint on the reconstructed elasticity modulus. 
\vspace{-0.25cm}
\section{Adversarial Learning of Regularizer}
\vspace{-0.15cm}
\label{sec:reg}
The core idea behind adversarial learning of regularizers is that a good regularizer $\mathbf{C}_{w}(\mathbf{x})$ should be able to distinguish the distribution of ground-truth elasticity images $\mathbb{P}_{r}$ from the distribution of the noisy images $\mathbb{P}_{n}$. It should be mentioned that the noisy elasticity images are reconstructed by maximum likelihood (ML) estimation in the unregularized optimization problem given the noisy force measurements $\mathbf{f}$ and forward operator $\mathbf{D(u^{m})}$; therefore, these images are corrupted with the correlated noise with covariance $\bm{\Gamma}$. The 1-Wasserstein distance involves the minimum path length to transport mass from one distribution to the other \cite{carola2}. WGANs approximate that mapping by training a convolutional neural network as regularizer by minimizing the following cost:
\vspace{-0.15cm}
\begin{equation}
\label{eq:10}
\mathbb{E}_{\mathbf{x}\sim \mathbb{P}_{r}}[\mathbf{C}_{w}(\mathbf{x})]-\mathbb{E}_{\mathbf{x}\sim \mathbb{P}_{n}}[\mathbf{C}_{w}(\mathbf{x})]+\mu \mathbb{E}[(\left \| \nabla _{\mathbf{x}}\mathbf{C}_{w}(\mathbf{x}) \right \|-1)_{+}^{2}]
\end{equation}
The first two terms aim to ensure that the learned network will be able to map the noisy image distribution to the ground-truth distribution. The last term is added for the stability of the critic network during training which constrains that the learned network is Lipschitz continuous with constant 1 and penalty coefficient $\mu$ \cite{carola}. Using the loss function in (\ref{eq:10}) and sampling ground-truth images $\mathbf{x}_{r}\sim \mathbb{P}_{r}$ and noisy ones $\mathbf{x}_{n}\sim \mathbb{P}_{n}$, we introduce an intermediate sample $\mathbf{x}_{i}$ for applying the Lipschitz constraint as $\mathbf{x}_{i}=\nu \mathbf{x}_{r}+(1-\nu)\mathbf{x}_{n}$, where $\nu \in U[0,1]$ is a uniformly-sampled scale in the range of [0,1]. By employing this intermediate sample, the loss function in each iteration of network training becomes:
\begin{equation}
\label{eq:12}
L_{w}=\mathbf{C}_{w}(\mathbf{x}_{r})-\mathbf{C}_{w}(\mathbf{x}_{n})+\mu [(\left \| \nabla _{\mathbf{x}_{i}}\mathbf{C}_{w}(\mathbf{x}_{i}) \right \|-1)_{+}^{2}]
\end{equation}
The training procedure of this adversarial regularizer is depicted in Fig \ref{fig:1}. It is worth noting that the minimization of the aforementioned loss function on image distributions allows some level of unsupervised learning, as it can involve the use of unpaired training. In particular, the  ground truth and noisy images can be independent (i.e., not necessarily paired) samples from $\mathbb{P}_{r}$ and $\mathbb{P}_{n}$, respectively. Once the network is trained, the learned explicit regularizer is plugged into (\ref{eq:9}) for updating the elasticity modulus $\mathbf{x}$. This reconstruction procedure is illustrated in Fig. \ref{fig:2}.
\begin{figure*}[t!]%[htb]
\begin{minipage}[b]{0.19\linewidth}
  \centering
  \centerline{\includegraphics[width=3.2cm]{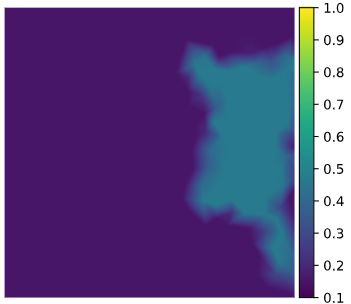}}
    \vspace{-0.6\baselineskip}
%  \vspace{1.5cm}
  \centerline{ \scriptsize{}}\medskip
\end{minipage}
\begin{minipage}[b]{.19\linewidth}
  \centering
  \centerline{\includegraphics[width=2.7cm]{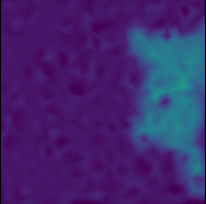}}
    \vspace{-0.6\baselineskip}
%  \vspace{1.5cm}
  \centerline{ \scriptsize{}}\medskip
\end{minipage}
%\hspace{1cm}
\begin{minipage}[b]{0.19\linewidth}
  \centering
  \centerline{\includegraphics[width=2.7cm]{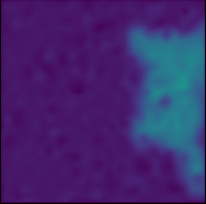}}
    \vspace{-0.6\baselineskip}
%  \vspace{1.5cm}
  \centerline{ \scriptsize{}}\medskip
\end{minipage}
%\par\vspace{+0.25\baselineskip}
\begin{minipage}[b]{0.19\linewidth}
  \centering
  \centerline{\includegraphics[width=2.7cm]{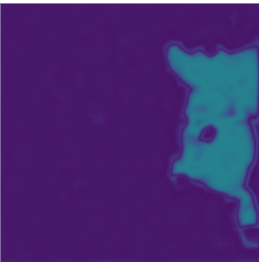}}
    \vspace{-0.6\baselineskip}
%  \vspace{1.5cm}
  \centerline{ \scriptsize{}}\medskip
\end{minipage}
%\hspace{1cm}
\begin{minipage}[b]{0.19\linewidth}
  \centering
  \centerline{\includegraphics[width=2.7cm]{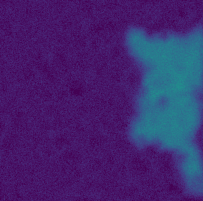}}
    \vspace{-0.6\baselineskip}
%  \vspace{1.5cm}
  \centerline{ \scriptsize{}}\medskip
\end{minipage}
%\hspace{1cm}
\par\vspace{-0.5\baselineskip}
\begin{minipage}[b]{0.19\linewidth}
  \centering
  \centerline{\includegraphics[width=3.3cm]{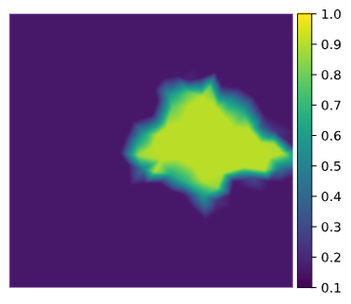}}
    \vspace{-0.5\baselineskip}
%  \vspace{1.5cm}
  \centerline{(a) \scriptsize{}}\medskip
\end{minipage}
\begin{minipage}[b]{.19\linewidth}
  \centering
  \centerline{\includegraphics[width=2.7cm]{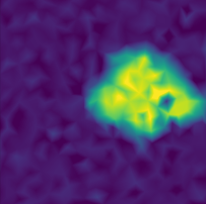}}
    \vspace{-0.3\baselineskip}
  %\vspace{1.5cm}
  \centerline{(b) \scriptsize{}}\medskip
\end{minipage}
%\hspace{1cm}
\begin{minipage}[b]{0.19\linewidth}
  \centering
  \centerline{\includegraphics[width=2.7cm]{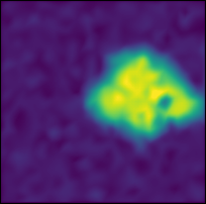}}
    \vspace{-0.3\baselineskip}
%  \vspace{1.5cm}
  \centerline{(c) \scriptsize{}}\medskip
\end{minipage}
%\par\vspace{+0.25\baselineskip}
\begin{minipage}[b]{0.19\linewidth}
  \centering
  \centerline{\includegraphics[width=2.7cm]{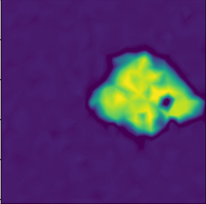}}
    \vspace{-0.3\baselineskip}
%  \vspace{1.5cm}
  \centerline{(d) \scriptsize{}}\medskip
\end{minipage}
%\hspace{1cm}
\begin{minipage}[b]{0.19\linewidth}
  \centering
  \centerline{\includegraphics[width=2.7cm]{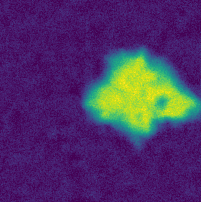}}
    \vspace{-0.3\baselineskip}
%  \vspace{1.5cm}
  \centerline{(e) \scriptsize{}}\medskip
\end{minipage}
%\hspace{1cm}
\vspace{-0.35cm}
\caption{(a) Ground-truth elasticity image. (b) Reconstructed elasticity image using unregularized optimization. (c) Reconstructed image with post-processing approach using UNet. (d) Reconstructed image using PnP approach and DnCNN. (e) Reconstructed elasticity image with the proposed adversarial learning-based regularization approach. The unit of the color bar is 100 KPa. }
\vspace{-0.5cm}
\label{fig:3}
\end{figure*}
\begin{figure}[h]%[htb]
\begin{minipage}[b]{0.8\linewidth}
  \centering
  \centerline{\includegraphics[width=6.5cm]{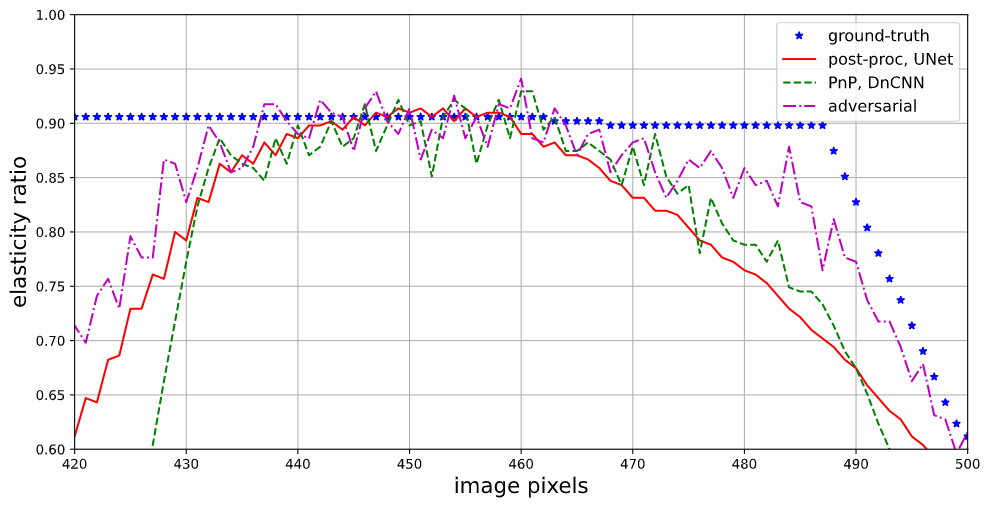}}
    \vspace{-0.4\baselineskip}
%  \vspace{1.5cm}
  \centerline{(a) \scriptsize{}}\medskip
\end{minipage}
\begin{minipage}[b]{.19\linewidth}
  \centering
  \centerline{\includegraphics[width=2.1cm]{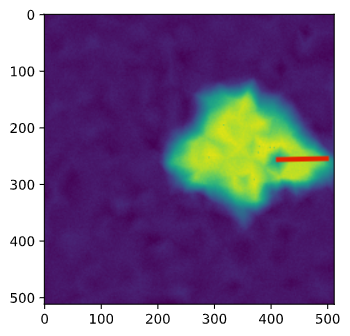}}
    \vspace{-0.4\baselineskip}
%  \vspace{1.5cm}
  \centerline{(b)\scriptsize{}}\medskip
\end{minipage}
%\hspace{1cm}
  \vspace{-0.85cm}
\caption{ (a) The cross section of reconstructed elasticity images using different approaches marked by the red line in (b). }
\label{fig:4}
\vspace{-0.5cm}
\end{figure}
\vspace{-0.2cm}
\section{Simulations and results}
\vspace{-0.15cm}
\vspace{-0.25cm}
\begin{table}[htb]
\centering\begin{tabular}{ | m{6.5em} | m{5cm}| } %| m{0.6cm} 
\hline
method & architecture \\ 
\hline
supervised reconstruction: Post-processing NN & UNet: 4 strided conv. (stride=2) + 4 transposed conv., Leaky ReLU and skip connection after each layer, Adam optimizer with lr=1e-3 \\
\hline
supervised reconstruction: PnP & DnCNN : 1 strided conv. (stride=1) + 10 (conv. layer,BN) + 1 conv., ReLU after each layer excep the last one, Adam optimizer with lr=1e-3 \\ 
\hline
unsupervised reconstruction: adversarial & 2 conv. + 4 strided conv. (stride=2) + 2 dense layers, Leaky ReLU after each conv. layers, RMSProp optimizer with lr=1e-4\\ 
\hline
\end{tabular}
\vspace{-0.25cm}
\caption{Details of implemented networks for elasticity reconstruction.}
\label{table1}
\vspace{-0.5cm}
\end{table}
To evaluate the performance of the proposed approach for solving the elasticity imaging problem, we train the adversarial regularizer network using true ground-truth elasticity images and observed measurements of force $\mathbf{f}$ and noisy deformations $\mathbf{u^{m}}$. We use a dataset of 541 B-mode images of the real lesion in breast tissue provided in \cite{dataset} to generate clean elasticity images (synthetic $\mathbf{x}$ maps). To this end, we generate normalized lesion elasticities in the range 0.3-0.8 KPa and normalized background elasticities in the range 0.1-0.15 KPa. With these choices, the ratio of lesion elasticity to the background elasticity falls in the range of 2-8, which represents experimental scenarios well. Moreover, the deformation images $\mathbf{u^{m}}$ are obtained for each elasticity image $\mathbf{x}$ by solving the forward model in (\ref{eq:1}) and adding multivariate Gaussian noise $\mathbf{n}$ \cite{narges2} resulting in $SNR=35dB$. While training the network, clean elasticity images with size $512\times512$ are fed into the network architecture and network parameters are learned using the loss function minimization process described in Section \ref{sec:reg}. with RMSProp optimizer. The details of the implemented network architecture for learning the adversarial regularizer is provided in Table \ref{table1}. Other network settings can be described as batch-size=16, num-epochs=100, Lipschitz regularizer scale $\mu=5$, adversarial regularizer scale $\lambda=10$, gradient step $\epsilon=0.7$, and the number of steps of gradient descent taken on the loss function is set to 100. These hyper-parameters are set by analyzing the reconstructed image quality.
For reconstruction performance comparison, we also implement two supervised learning approaches: post-processing neural network (NN) \cite{postprocessing} using a UNet architecture and PnP paradigm using a DnCNN architecture \cite{tum}. For generating noisy elasticity images, we map the noisy displacement fields $\mathbf{u^{m}}$ to the image domain by solving the unregularized inverse problem consisting of a data-fidelity term and a positivity constraint.
 The simulation results for reconstruction of elasticity images using these different approaches are presented in Fig. \ref{fig:3}. The UNet denoiser blurs the image to remove the artifacts while the other learning-based methods preserve the edges more efficiently. The cross-section details of each reconstruction presented in Fig. \ref{fig:4} indicate that our proposed adversarial approach has better reconstruction performance compared to the other methods. For computation time comparison, since PnP with DnCNN architecture uses an iterative scheme for image reconstruction, its computation time is higher than the post-processing method in the test time. Likewise, our adversarial approach requires more computation time for network training using the distribution-based loss and also image reconstruction in the test time.  \cmmnt{For this unsupervised learning method, the train network is used as the regularizer (prior) which constrain reconstructed image as the solution of inverse problem to lies on this learned manifold.}
\vspace{-0.3cm}
\section{Conclusion}
\vspace{-0.2cm}
This article proposes a new learning-based approach for ultrasound elasticity imaging by combining physical modeling and adversarial regularizer learning. The powerful image regularizer is trained based on the Wasserstein distance loss for estimating the distribution of latent elasticity images which allows some level of unsupervised learning as well. Then, the learned explicit regularizer is plugged into the optimization task as the prior information to mitigate the noisy and corrupted measurements. Finally, the resulting minimization problem composed of a data-fidelity term and the learned adversarial regularization term is solved using gradient descent to reconstruct the estimates of latent elasticity images. Our preliminary simulation results demonstrate the effectiveness of the proposed method in elasticity image reconstruction in terms of robustness and accuracy.
\bibliographystyle{IEEEbib}
\bibliography{Mybib}

\begin{thebibliography}{10}

\bibitem{Marvin}
M.~M. Doyley,
\newblock ``Model-based elastography: a survey of approaches to the inverse
  elasticity problem.,''
\newblock {\em Phys. in med. and biol.}, vol. 57 3, pp. R35--73, 2012.

\bibitem{narges2}
N.~Mohammadi, M.~Doyley, and M.~Cetin,
\newblock ``Ultrasound elasticity imaging using physics-based models and
  learning-based plug-and-play priors,''
\newblock {\em ICASSP}, pp. 1165--1169, 2021.

\bibitem{salman2}
S.~Mohamadi and H.~Amindavar,
\newblock ``Deep bayesian active learning, a brief survey on recent advances,''
\newblock {\em arXiv preprint arXiv:2012.08044}, 2020.

\bibitem{E2E}
B.~Ni and H.~Gao,
\newblock ``A deep learning approach to the inverse problem of modulus
  identification in elasticity,''
\newblock {\em MRS Bulletin}, p. 1–7, 2020.

\bibitem{E2E2}
S.~Wu, Z.~Gao, and et~al.,
\newblock ``Direct reconstruction of ultrasound elastography using an
  end-to-end deep neural network,''
\newblock in {\em MICCAI}, 2018.

\bibitem{E2E3}
R.~R. {Wildeboer}, R.~J.~G. v.~{Sloun}, and et~al.,
\newblock ``Synthetic elastography from {B}-mode ultrasound through deep
  learning,''
\newblock in {\em IUS}, 2019, pp. 108--110.

\bibitem{tehrani}
A.~K.~Z. Tehrani and H.~Rivaz,
\newblock ``{MPWC-Net++}: evolution of optical flow pyramidal convolutional
  neural network for ultrasound elastography,''
\newblock in {\em SPIE}, 2021.

\bibitem{mahsa1}
M.~Mozaffari and Y.~Yilmaz,
\newblock ``Online anomaly detection in multivariate settings,''
\newblock {\em MLSP}, pp. 1--6, 2019.

\bibitem{willet}
G.~{Ongie}, A.~{Jalal}, C.~A. {Metzler}, R.~G. {Baraniuk}, A.~G. {Dimakis}, and
  R.~{Willett},
\newblock ``Deep learning techniques for inverse problems in imaging,''
\newblock {\em IEEE J. on Sel. Areas in Inf. Theory}, vol. 1, no. 1, pp.
  39--56, 2020.

\bibitem{PGD}
M.~Mardani, Q.~Sun, and et~al.,
\newblock ``Neural proximal gradient descent for compressive imaging,''
\newblock in {\em NeurIPS}, 2018.

\bibitem{selfsuper}
O.~Senouf, S.~Vedula, and et~al.,
\newblock ``Self-supervised learning of inverse problem solvers in medical
  imaging,''
\newblock {\em ArXiv}, vol. abs/1905.09325, 2019.

\bibitem{ali}
M.~A. Vosoughi and S.~Köse,
\newblock ``Combined distinguishers to enhance the accuracy and success of side
  channel analysis,''
\newblock {\em ISCAS}, pp. 1--5, 2019.

\bibitem{PINN}
E.~Haghighat, M.~Raissi, and et~al.,
\newblock ``A deep learning framework for solution and discovery in solid
  mechanics: linear elasticity,''
\newblock {\em ArXiv}, vol. abs/2003.02751, 2020.

\bibitem{GAN}
J.~E. Warner, J.~Cuevas, and et~al.,
\newblock ``Inverse estimation of elastic modulus using physics-informed
  generative adversarial networks,''
\newblock {\em ArXiv}, vol. abs/2006.05791, 2020.

\bibitem{Moji}
M.~Heydari and Z.~Duan,
\newblock ``Don’t look back: An online beat tracking method using {RNN} and
  enhanced particle filtering,''
\newblock {\em ICASSP}, pp. 236--240, 2021.

\bibitem{salman1}
S.~Mohamadi, D.~Adjeroh, B.~Behi, and H.~Amindavar,
\newblock ``A new framework for spatial modeling and synthesis of genomic
  sequences,''
\newblock {\em BIBM}, pp. 2221--2226, 2020.

\bibitem{pnp}
Y.~{Sun}, B.~{Wohlberg}, and U.~S. {Kamilov},
\newblock ``An online plug-and-play algorithm for regularized image
  reconstruction,''
\newblock {\em IEEE Trans. on Comput. Imag.}, vol. 5, no. 3, pp. 395--408,
  2019.

\bibitem{RED}
Y.~Romano, M.~Elad, and P.~Milanfar,
\newblock ``The little engine that could: Regularization by denoising
  ({RED}),''
\newblock {\em SIAM J. on Imag. Sci.}, vol. 10, no. 4, pp. 1804--1844, 2017.

\bibitem{primaldual}
J.~Adler and O.~{\"O}ktem,
\newblock ``Learned primal-dual reconstruction,''
\newblock {\em IEEE Trans. on Med. Imag.}, vol. 37, pp. 1322--1332, 2018.

\bibitem{GANprior}
D.~V. Patel and A.~A. Oberai,
\newblock ``Quantifying uncertainty with {GAN}-based priors,''
\newblock {\em ArXiv}, vol. abs/2003.12597, 2020.

\bibitem{WGAN}
M.Arjovsky, S.~Chintala, and L.~Bottou,
\newblock ``Wasserstein generative adversarial networks,''
\newblock in {\em ICML}, 2017.

\bibitem{morteza2}
K.~{Lei}, M.~{Mardani}, and et~al.,
\newblock ``Wasserstein {GAN}s for {MR} imaging: From paired to unpaired
  training,''
\newblock {\em IEEE Trans. on Med. Imag.}, vol. 40, no. 1, pp. 105--115, 2021.

\bibitem{carola}
S.~Lunz, O.~{\"O}ktem, and C.~Sch{\"o}nlieb,
\newblock ``Adversarial regularizers in inverse problems,''
\newblock in {\em NeurIPS}, 2018.

\bibitem{carola2}
S.~Arridge, P.~Maass, O.~{\"O}ktem, and C.~Sch{\"o}nlieb,
\newblock ``Solving inverse problems using data-driven models,''
\newblock {\em Acta Numerica}, vol. 28, pp. 1 -- 174, 2019.

\bibitem{dataset}
W.~Al-Dhabyani, M.~Gomaa, H.~Khaled, and A.~Fahmy,
\newblock ``Dataset of breast ultrasound images,''
\newblock {\em Data in Brief}, vol. 28, pp. 104863, 2020.

\bibitem{postprocessing}
K.~H. {Jin}, M.~T. {McCann}, E.~{Froustey}, and M.~{Unser},
\newblock ``Deep convolutional neural network for inverse problems in
  imaging,''
\newblock {\em IEEE Trans. on Image Proc.}, vol. 26, no. 9, pp. 4509--4522,
  2017.

\bibitem{tum}
T.~Meinhardt, M.~M{\"o}ller, C.~Hazirbas, and D.~Cremers,
\newblock ``Learning proximal operators: Using denoising networks for
  regularizing inverse imaging problems,''
\newblock {\em ICCV}, pp. 1799--1808, 2017.

\end{thebibliography}
%\printbibliography
\end{document}